\documentclass[twocolumn]{aastex6}
\usepackage{amsmath,amssymb,bm, graphicx,xspace,multirow,threeparttable}
\usepackage{epsfig}
\usepackage{color}
\RequirePackage{lineno}
\usepackage{txfonts}
\usepackage{float}

\usepackage{color}

\def\deg{\hbox{$^\circ$}}

\def\aap{Astronomy and Astrophysics}

\def\apjl{ApJL}

\def\apjs{ApJS}

\begin{document}

\title{Measuring the Mass of Missing Baryons in the Halo of Andromeda Galaxy with Gamma-Ray Observations}
\author{Yi Zhang$^{1,2}$, Ruo-Yu Liu$^{1,2,*}$, Hui Li$^{3,4,\dag}$, Shi Shao$^5$, Huirong Yan$^{6,7}$ and Xiang-Yu Wang$^{1,2}$}

\affil{$^1$School of Astronomy and Space Science, Xianlin Road 163, Nanjing University, Nanjing
210023, China; *Email:\textcolor{blue}{ryliu@nju.edu.cn}\\
$^2$Key laboratory of Modern Astronomy and Astrophysics (Nanjing University), Ministry of Education, Nanjing 210023, People's Republic of China\\
$^3$Department of Physics and Kavli Institute for Astrophysics and Space Research, Massachusetts Institute of Technology, Cambridge, MA 02139, USA\\
$^4$Department of Astronomy, Columbia University, 550 West 120th Street, New York, NY, 10027, USA\\
$^5$Institute for Computational Cosmology, Department of Physics, Durham University, South Road Durham DH1 3LE, UK\\
$^6$Deutsches Elektronen Synchrotron (DESY), Platanenallee 6, D-15738 Zeuthen, Germany\\
$^7$Institut f\"ur Physik und Astronomie, Universit\"at Potsdam, D-14476 Potsdam, Germany\\
$^\dag$ NHFP Hubble Fellows}

\begin{abstract}
It has been suggested that a huge amount of baryons resides in the circumgalactic medium (CGM) extending out to the virial radii of galaxies. In this work we attempt to measure the baryon mass in CGM with a novel method based on the gamma-ray observations of the extended halo of the Andromeda Galaxy Messier 31 (M31). Since cosmic-ray particles generated inside the galaxy will eventually escape to the CGM, they will produce gamma-ray emission via the proton-proton collision with CGM and produce gamma rays. Different from some traditional measurements which are sensitive only to certain metallic ions in specific temperature range, the hadronic gamma-ray flux is sensitive to baryonic gases in all phases and does not rely on the metallicity in the halo, hence the measured gamma-ray flux from the galaxy's halo can be used to constrain the mass of CGM. By dealing with the cosmic-ray transport in the halo and calculating the hadronic gamma-ray intensity, we find that the total baryon mass contained within the virial radius is less than $(1.4-5)\times 10^{10}M_\odot$ according to the gamma-ray intensity obtained with a model-dependent analysis. It implies that the CGM of Andromeda Galaxy may not account for more than 30\% of the missing baryons, but the result is subject to uncertainties from the gamma-ray intensity upper limit, diffusion coefficient of the CRs in the halo as well as the stellar mass and dark matter halo mass of the galaxy. This method will become more constraining provided better understandings on these issues and more sensitive gamma-ray telescopes in the future.
\end{abstract}

\maketitle

\section{Introduction}\label{sec:intro}
While baryons are measured to account for 16\% of the total mass of the universe according to the observations of the cosmic microwave background \cite{Planck18}, more than half of this ordinary matter that makes up almost everything familiar to us is yet to be localised \citep{Persic92, Fukugita98}. Decades of efforts have been dedicated to the hunt for missing baryons, and 
various studies have pointed to the existence of a gaseous halo extending out to at least 50\,kpc around Milky Way and external galaxies with stellar luminosity similar to that of Milky Way, containing a total gas mass of a few times $10^{9}M_\odot$ within 50\,kpc \citep{Anderson10, Dai12, Fang13, Miller13, Miller15,LiJT17}. The gas density of the CGM $n(r)$ as a function of galactocentric radius $r$ could be depicted by the so-called $\beta$ model\citep[][or a modified $\beta$ model, see \citealt{Bogdan13}]{Cavaliere76}, i.e., $n(r)\propto n_0(1+(r/r_c)^2)^{-3\beta/2}$ where $n_0$ is a normalization factor, $r_c$ is the core radius with the typical value $\lesssim 5\,$kpc, and $\beta$ determines the slope of the distribution. The value of $\beta$ within $50\,$kpc is found to be around 0.5 in many Milky Way-like galaxies \citep{LiJT18, Bregman18}. The total CGM mass within the virial radius is mainly determined by the gas distribution at larger radius. 
If the slope for gas distribution within $50\,$kpc holds out to a much larger radius \citep{LiJT18} or even the virial radius, the CGM mass in the halo would be significantly less than that needed to account for the missing baryons \citep{Miller15, Bregman18}. On the other hand, if a flattening is present in the density distribution at comparatively small radius with, e.g., $\beta\lesssim 0.3$, the entire halo could contain all the missing baryon\citep{Gupta12,Faerman17, Das19}. Therefore, the key to evaluate the total missing baryon in CGM is to determine the density distribution slope beyond 50\,kpc. 

The halo of our Milky Way Galaxy has the most abundant observational data among all the galaxies and it has been extensively studied by various groups. However, the observable signals from different galactocentric radius of the halo all add up together along our line of sights, making it difficult to reveal its true gas distribution. The Andromeda Galaxy, the nearest sibling of our Milky Way Galaxy, is probably the best target to study this issue. Unlike the halo of our Galaxy, we can observe Andromeda's halo from the side view and hence are, in principle, able to measure the baryon density as a function of the galactocentric radius. Moreover, Andromeda's halo is more detectable than other Milky Way-like external galaxies due to its proximity, provided that their intrinsic emissivities are similar. Indeed, evidence of a massive gaseous halo around Andromeda has already been discovered \citep{Grcevich09, Lehner15, Lehner20}. The project Absorption Maps In the Gas of Andromeda (AMIGA) researches the physical conditions and metals in the CGM of M31 through the measurements of ultraviolet absorption along 43 QSO sightlines. The inferred baryon mass in cold and warm gas is about $4\times 10^{10}M_\odot$ within the virial radius assuming a 0.3 solar metallicity for the CGM \citep{Lehner20}. {\citet{Lehner20} also reported that the column densities of ions which indicate warm gases show a comparatively shallower decrease as the function of the projected distance to the galaxy beyond $\sim 100\,$kpc than those at small distance.} Nevertheless, many traditional methods of probing CGM such as those with UV/X-ray emission lines or absorption lines do not work well beyond 50\,kpc because of the weak emission of the tenuous gas and the high Galactic foreground noise, as well as the large uncertainty from the metallicity of the CGM.


The Andromeda Galaxy has been detected in GeV gamma-rays by the {\it Fermi} Large Area Telescope (LAT)  \citep{Fermi17_M31_excess, Fermi10_localgroup}. The gamma-ray signal may originate from multiple possible processes. In addition to the dark matter annihilation and unresolved population of millisecond pulsars \citep{Pshirkov16, Fermi17_M31_excess, Eckner18, McDaniel18, Fragione19, McDaniel19}, part of the gamma-ray emission may arise from the proton-proton collisions between cosmic-ray (CR) protons accelerated in the galaxy and the interstellar medium. Referring to the situation in our Galaxy, CRs typically leave the galaxy quickly on a timescale of $\sim10$\,Myr, losing only a minor fraction of their energies, and diffuse into the extended halo of the galaxy. Those wandering CRs will inevitably interact with CGM and produce gamma-ray emission via the proton--proton collision \citep{Feldmann13, Taylor14, Kalashev16, Liu19_CGM, Biswas19, Jana20}, same as that responsible for gamma rays from the galactic plane of our Galaxy. \citet{Feldmann13} pointed out that the gamma-ray flux generated via this process in the halo of our Galaxy is approximately proportional to the amount of the gas residing in the halo and contributes to the isotropic gamma-ray background. Therefore, the gamma-ray signal from the halo contains the information of the total mass of the baryonic matter therein. \citet{Liu19_CGM} found that the measurement of the isotropic gamma-ray background may constrain the baryon mass within the virial radius of our Galaxy to be lower than $3\times 10^{10}M_\odot$ in the fiducial model. Here we extend this method to the extended halo of the Andromeda Galaxy by considering the CR diffusion and the hadronic emission in the CGM, with the advanced gamma-ray analysis managing to identify the emission of the galaxy's halo from the foreground/background, as introduced below.


The diffuse gamma-ray emission was detected recently extending out to 200\,kpc from the Andromeda galaxy \citep{Karwin19}. The intensity for $\lesssim 10\,$GeV gamma-rays is at the level of $\sim 10^{-7}\,\rm GeVcm^{-2}s^{-1}sr^{-1}$, while no significant emission is detected above 10\,GeV, which poses a quite constraining upper limit of 68\% confidence level (C.L.) $\lesssim 10^{-8}\,\rm GeVcm^{-2}s^{-1}sr^{-1}$ at $\approx 35\,$GeV assuming the spectrum to be a power-law function with exponential cutoff (PLEXP). Besides, a model-independent upper limit is given to be $\lesssim 2\times 10^{-7}\,\rm GeVcm^{-2}s^{-1}sr^{-1}$ at $\approx 35$\,GeV by freeing the spectral index in each energy bin in the analysis. In addition to CR interactions, other sources, such as termination shocks of a galactic wind \citep{Zirakashvili06} and dark matter annihilation \citep{Karwin20}, could also contribute to the diffuse gamma-ray emission of the halo, and thus the measured gamma-ray flux or upper limit should be regarded as a conservative upper bound for CR-generated gamma-ray emission. 
This would enable us to constrain the gas distribution beyond 50\,kpc by comparing the measured intensity to the predicted gamma-ray intensity from the $pp$ collision under different assumptions of the slope (i.e., $\beta$), and test whether a flattening of the gas distribution is allowed by the gamma-ray data.

The rest of the paper is organized as follows: in Section 2, we model the CR distribution in the halo of M31 and the gamma-ray production via the $pp$ collision. We compare the predicted gamma-ray intensity with the measured one and derive the constraints on the gas distribution beyond 50\,kpc. We discuss the result obtained in this work in Section 3 and summarize the paper in Section 4.

\section{Cosmic-Ray distribution in the halo of M31 and the pionic gamma-ray flux }

\begin{figure}[htbp]
\centering
\includegraphics[width=0.9\columnwidth]{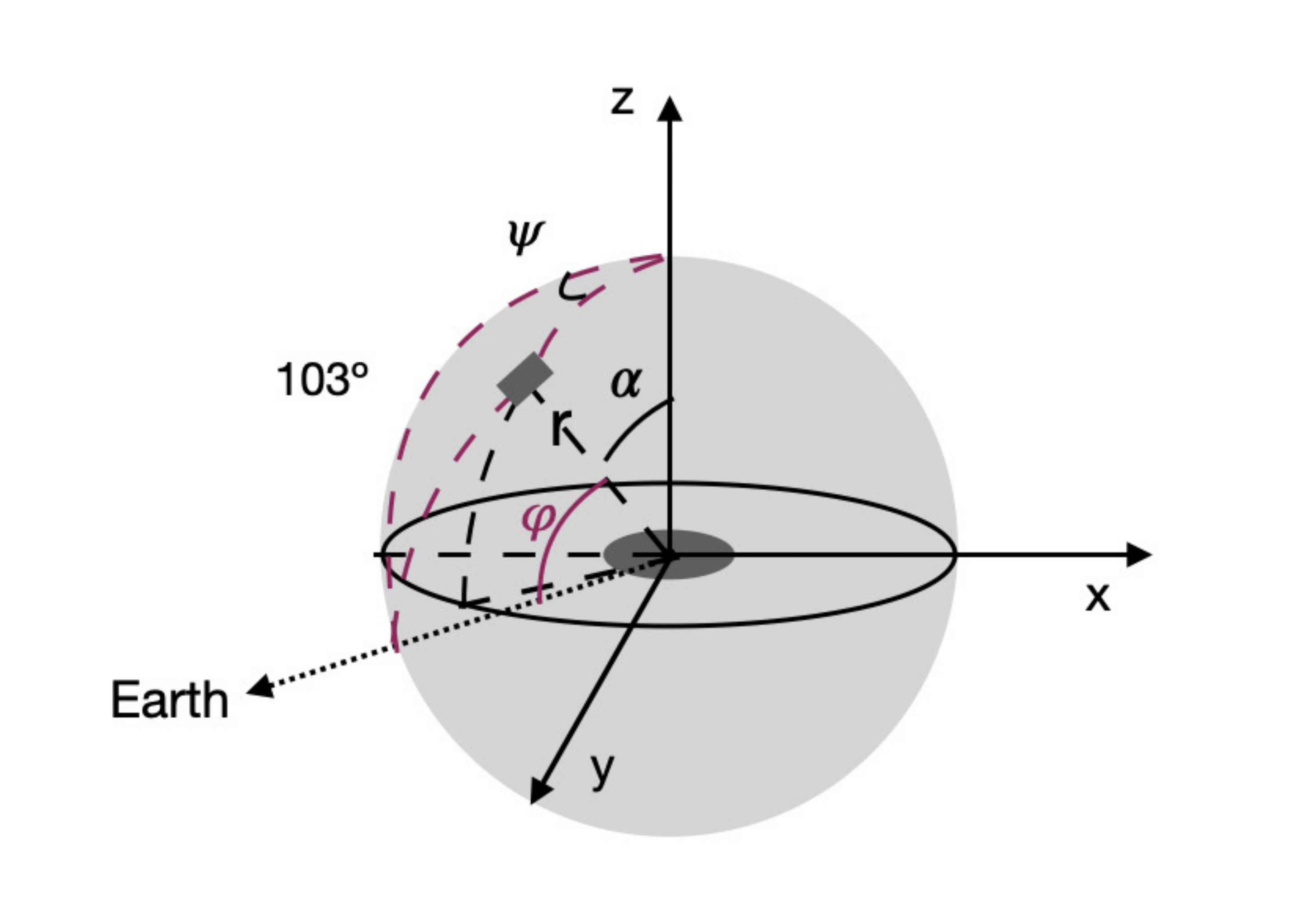}
\includegraphics[width=0.9\columnwidth]{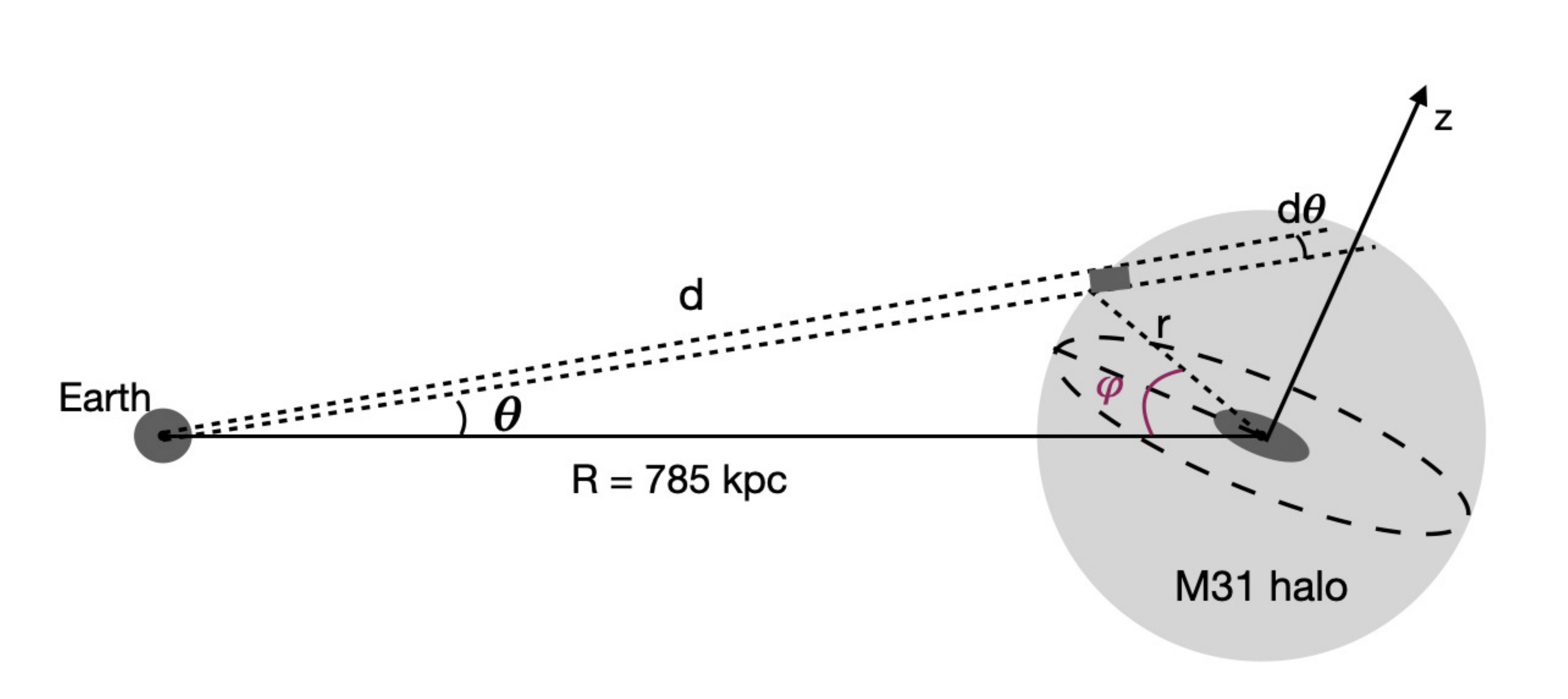}
\caption{The diagrammatic sketch of M31 coordinate system and the geometry between M31 and Earth. The M31 coordinate system is symmetric with respect to the axis perpendicular with disk plane. The angle between the north pole and Earth is $103^\circ$. The shaded rectangle is the element volume in the integration with the coordinate $(r,\alpha,\psi)$. $\phi$ is the open angle between the earth and the element volume. The spherical triangular highlighted by purple dashes  is used to calculate $\phi$. $d$ is the distance between the element volume and Earth and $\theta$ is the viewing angle, which can be calculated from $r,\phi,R$ using simple geometric relationship. }
\label{fig:m31}
\end{figure}

The transport equation of CRs injected from a single point source is \citep{Berezinsky06b}
\begin{equation}
    \frac{\partial n}{\partial t}=D(E)\nabla _{\pmb{r}}^2 n -v_w\nabla_{\pmb{r}}n + \frac{\partial [b(\pmb{r},E,t)]}{\partial E} +Q(E,t)\delta^3(\pmb{r}-\pmb{r}_s),
\end{equation}
where $n(\pmb{r},E,t)$ is the differential density of CRs at time $t$ and position $\pmb{r}(r,\alpha,\psi)$.
We use a spherical coordinate system with the original point at the center of M31 and the $z$-axis being the normal of the disk (see Fig.\ref{fig:m31}). The first term on the r.h.s of the equation describes the diffusion of CRs caused by magnetic field turbulence with the diffusion coefficient $D(E)$. The most constraining gamma-ray data is at 35\,GeV, so the most relevant proton energy is at 350\,GeV since the energy of gamma-ray photon generated by the $pp$ collision generally carries 10\% of the parent proton's energy. We thus denote the diffusion coefficient at 350\,GeV by $D_0$ and assume the form of the diffusion coefficient to be $D(E)=D_0(E/350{\rm GeV})^{1/3}$. It should be noted that the index of $D$ is not necessary to be $1/3$ or Kolmogorov-like \citep{Yan02}, but it would not influence our result significantly because the diffusion coefficient at 350\,GeV is mainly dependent on the value of $D_0$. The value of $D_0$ depends on the property (such as strength) of the turbulence. It is important for the CR spatial distribution in the halo and will be discussed in the later section. 

The second term describes the advection of CRs caused by large-scale galactic wind with a speed $v_w$. Currently there is no evidence for a large-scale galactic wind in the halo of M31. Note that \citet{Pshirkov16} found indication of bubble-like gamma-ray structure located perpendicular to the M31 disc with height $6-7.5$\,kpc, resembling the ``Fermi bubble'' \citep{Su10, Fermi14}. This kind of bubbles might represent an outflow driven by the past activity of the supermassive black hole in the center of the galaxy. While the outflow could advect CRs to larger radius, it is limited in the inner halo and hence we here simply assume $v_w=0$ in the later calculation.
The third term represents the energy loss of CRs due to the $pp$ collisions in CGM, which is given by $b(E,\pmb{r})=6\times 10^{-7}(E/10^{12}{\rm eV})(n(r)/10^{-3}\rm cm^{-3})\, \rm eV~s^{-1}$ \citep{Liu19_CGM}. 

\begin{figure}[htbp]
\centering
\includegraphics[width=0.9\columnwidth]{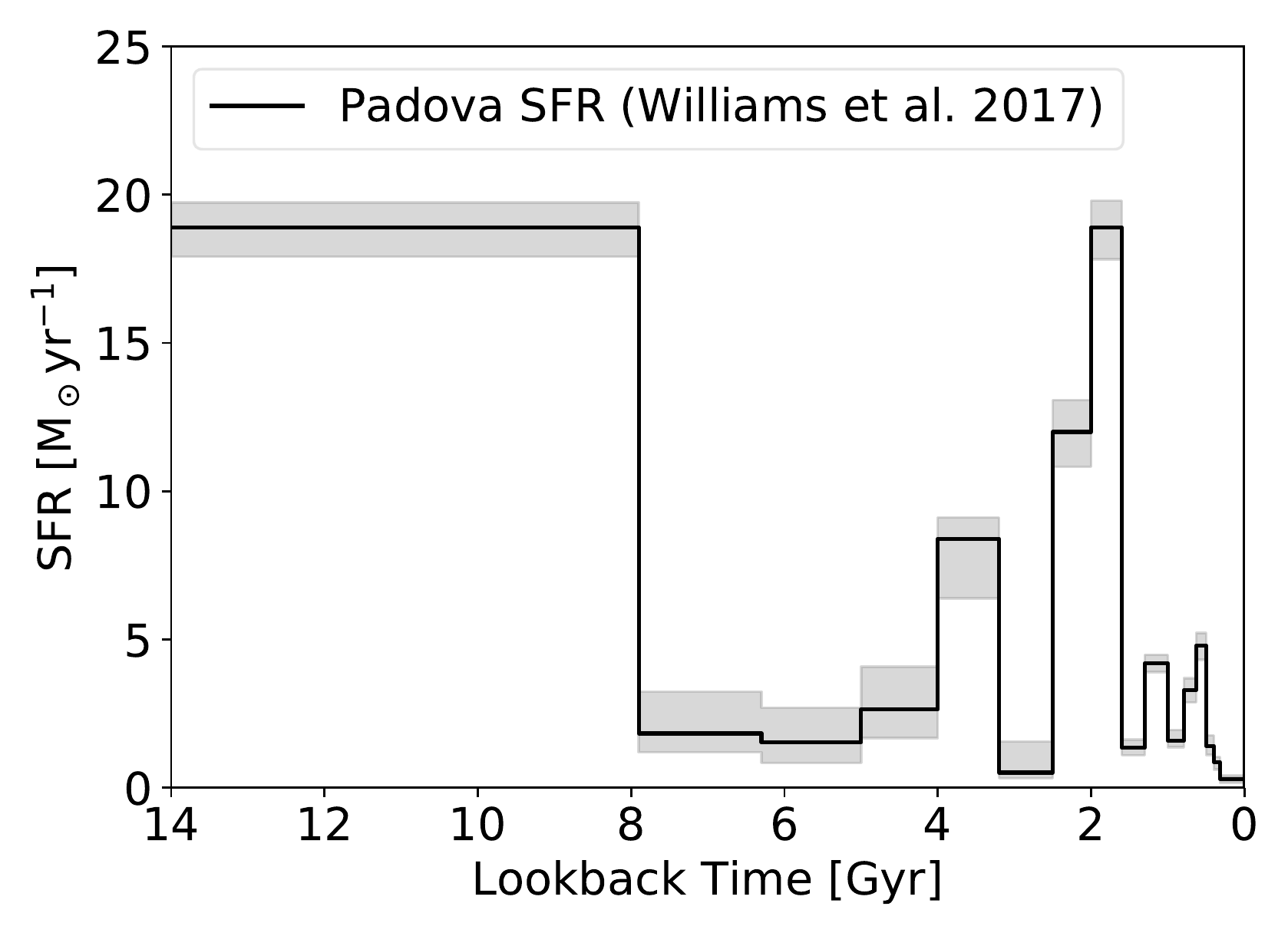}
\caption{Star formation history for M31 from present to 14 Gyr ago from Padova stellar evolution models \citep{Williams17}. The gray band shows the error region.}
\label{fig:sfh}
\end{figure}

The last term in the r.h.s. describes the injection of CRs from a point source located at $\pmb{r}_s$. $Q(E,t)$ can be decoupled into a term related to the CR injection history $S(t)$ and a term describing the CR injection spectrum $Q_0(E)$, i.e., $Q(E,t)=S(t)Q_0(E)$. By doing so, we implicitly assume that the shape of the injection spectrum does not evolve with time. The CR injection rate is generally proportional to the star formation rate \citep{Kennicutt98}. We here employ the star formation history of M31 derived by \citet{Williams17}, where four sets of the stellar evolution model are considered backtracking the star formation history of M31 to 14\,Gyr ago. All the models show a common feature that most of the star formation occurred prior to $8\,$Gyr ago. Here we refer to the Padova model, as is shown in Fig.\ref{fig:sfh}, and normalize the star-formation history with the present-day star formation rate ($\rm SFR_{\rm M31}$) so that we have $S(t=0)=1$.

Following \citet{Liu19_CGM}, we approximate the galactic plane as a disk of negligible thickness (i.e., $\pmb{r}_s=(r_s, \pi/2, \psi)$) and CRs are uniformly injected from the disk. 
The injection spectrum is assumed to be in the form of a power-law function, i.e., $Q_0(E)=N_0(E/1\,{\rm GeV})^{-s}$, where $N_0$ is the normalization factor and $s$ is the injection spectral index of CRs. Given the present CR luminosity in M31 $L_{\rm CR,0}$, we have $\pi R_{\rm gal}^2 \int^\infty_{1\,\rm GeV}EQ_0(E)dE=L_{\rm CR,0}$  with the radius $R_{\rm gal}=27\,$kpc being the radius of galaxy disk of M31. $L_{\rm CR,0}$ can be normalized to the CR luminosity of our Galaxy $L_{\rm CR, MW}=10^{41}\rm erg\,s^{-1}$  by $L_{CR,0}=L_{\rm CR, MW}\frac{\rm SFR_{\rm M31}}{\rm SFR_{\rm MW}}$. 
The measured gamma-ray spectral index of M31 galaxy is $2.4\pm 0.1$ \citep{Fermi10_localgroup}. We assume that the $pp$ collision dominates the gamma-ray emission at several tens GeV and the measured gamma-ray spectral slope represents the spectral index of the steady-state CR spectrum in the galaxy, since the gamma-ray spectrum from $pp$ collision generally follows the spectrum of the parent protons in the GeV energy range. Note that the energy-dependent escape of CRs from the galaxy leads to a softening in the steady-state spectrum with respect to the injection spectrum, with the spectral index being modified to $s+\Delta$ where $\Delta$ is the slope of the diffusion coefficient. In our Galaxy, $\Delta\simeq 1/3$ is inferred from the measurement of the primary-to-secondary CR ratio \citep{Aguilar16}. If the ISM in M31 has the similar property to that of our Galaxy, we may infer $s\gtrsim 2$ for the CR injection spectrum in M31, which is also consistent with the prediction of the canonical diffusive shock acceleration theory \citep{Bell78, Blandford87}. In the following calculation, we consider $s$ in the range of $2.0-2.2$.

\begin{figure}[htbp]
\centering
\includegraphics[width=0.9\columnwidth]{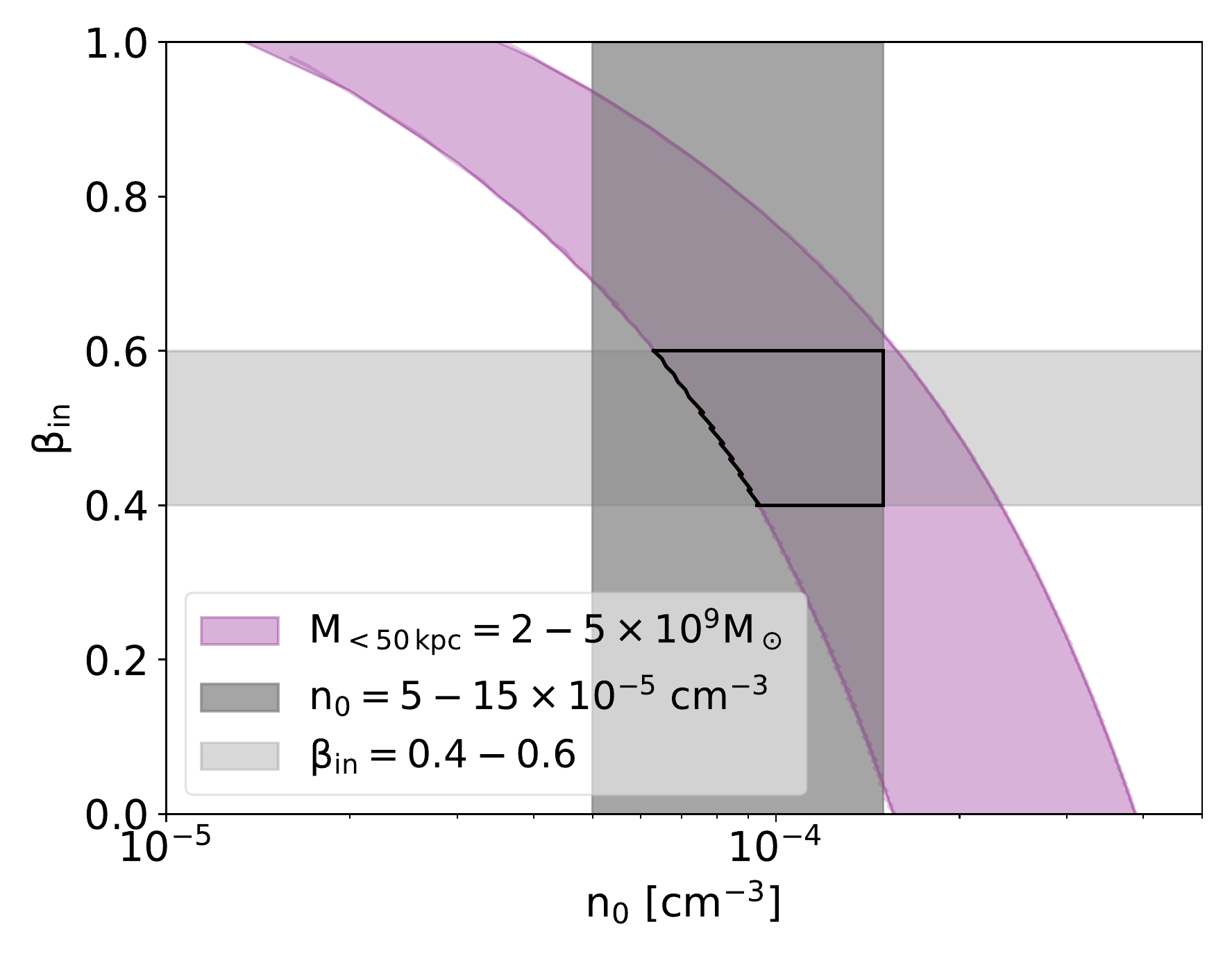}
\caption{The parameters of gas density profile within 50\,kpc. The gray and purple region marks different constraints on parameters from observation. The darkest region with black border marks the final employed parameter region. }
\label{gasin}
\end{figure}

An analytical formulae for the distribution of CR injected from a point source at $\pmb{r}_s$ can be given by \citep{Berezinsky06b}
\begin{equation}
\begin{split}
    n(\pmb{r},E,t;\pmb{r}_s)&=\frac{\pi^{3/2}}{(2\pi)^3}\int_{t_s}^t dt'Q(\mathcal{E}',t')\frac{\exp\left[-(\pmb{r}-\pmb{r}_s)^2/4\lambda(E,t')\right]}{\lambda(E,t')^{3/2}}\\
    &\times \exp\left[\int_{t'}^t dt''\frac{\partial b(\mathcal{E}'',t)}{\partial \mathcal{E}''}\right],
\end{split}
\end{equation}
where $\lambda(E,t')=\int_{t'}^t D(\mathcal{E}'')dt''$ implies the diffusion distance of a proton, $\mathcal{E}'(\mathcal{E}'')$ is the energy of a proton at time $t'(t'')$ and this proton has energy $E$ at present. We assume the CR sources homogeneously distributed throughout the disk and integrate over the contribution of each point sources in the disk to obtain the present-time ($t=0$) CR density at $\pmb{r}$ by
\begin{equation}
    N(\pmb{r},E)=\int n(t=0,\pmb{r},E;\pmb{r}_s)r_sdr_sd\psi
\end{equation}
After obtaining CR distribution, the gamma-ray emissivity can be calculated by \citep{Kelner06}, 
\begin{equation}
 J_\gamma(E_\gamma,\pmb{r})\equiv \frac{dN_\gamma}{dE_\gamma dt} =c n_g(\pmb{r}) \int_{E_\gamma}^\infty \sigma_{pp} N(\pmb{r},E)F_\gamma(\frac{E_\gamma}{E},E)\frac{dE}{E},
\end{equation}
where $F_\gamma$ is the spectrum of the secondary gamma-ray in a single collision, $n_g(r)$ is the gas density distribution in the halo. The gas density distribution within 50\,kpc is more or less known according to previous studies. The gas density of the halos of our Galaxy and Andromeda at a galactocentric radius $\backsimeq 50\,$kpc is measured to be $\approx 10^{-4}\rm cm^{-3}$ through modelling the ram-pressure on the clouds and satellite galaxies \citep{Salem15, Grcevich09}. We consider $n_0\equiv n(50{\rm kpc})=10^{-4}\rm cm^{-3}$ and $\beta_{\rm in}=0.5$ as the benchmark parameters. To include the uncertainties of parameters, we vary $\beta_{\rm in}$ in the range of $0.4-0.6$ and $n_0$ in the range $(0.5-1.5)\times 10^{-4}\,\rm cm^{-3}$. Additionally, following \citet{Anderson13, Bregman18} and etc, we limit the total CGM mass within 50\,kpc in the range of $(2-5)\times 10^9M_\odot$ by choosing appropriate combinations of $\beta_{\rm in}$ and the normalisation density $n_0$ (see Fig.~\ref{gasin}). On the other hand, for the gas distribution beyond 50\,kpc, a possible flattening or steepening in the distribution is taken into account. In other words, for $r\gg r_c$, we have the gas density distribution
\begin{equation}
n(r)=n_0\left\{
\begin{array}{ll}
\left(\frac{r}{50\rm kpc}\right)^{-3\beta_{\rm in}}, \quad r<50\,{\rm kpc} \\
\left(\frac{r}{50\rm kpc}\right)^{-3\beta_{\rm out}}, \quad r\geq 50\,{\rm kpc} \\
\end{array}
\right.
\end{equation}

M31 is approximately 785\,kpc away from Earth and the observation of its gamma-ray halo radius extends out to 200\,kpc from the galactic center. Denoting the coordinate of Earth by $\pmb{r}_E$, the distance between Earth and certain point in the halo $\pmb{r}$ can vary markedly from place to place. We therefore need to consider the geometry of the system to calculate the distance $|\pmb{r}-\pmb{r}_E|$ and the gamma-ray flux received at Earth. As shown in the top panel of Fig.~\ref{fig:m31}, for certain point (marked as the gray rectangular) in the halo, we set its coordinate as $(r,\alpha,\psi)$. Utilizing the spherical triangle highlighted with purple dashed curve, we find out the open angle $\phi$ between Earth and the point as $\cos\phi=\cos103\deg \cos\alpha+\sin 103\deg \sin\alpha \cos\psi$, and obtain $|\pmb{r}-\pmb{r}_E|$ by $\sqrt{(\pmb{r}-\pmb{r}_E)^2}=\sqrt{r^2+r_E^2-2rr_{\rm E}\cos\phi}$.
 Now we can calculate the average gamma-ray flux intensity  within an angular radius $\theta_c$ with respect to the center of M31 viewed from Earth in the celestial plane by
\begin{equation}
    \Phi_\gamma(E_\gamma)=\frac{1}{2\pi(\cos0.4^\circ -\cos\theta_c)}\int \frac{J_\gamma(E_\gamma,\pmb{r})}{4\pi (\pmb{r}-\pmb{r}_E)^2}dV.
\end{equation}
with $dV=r^2\sin\alpha d\alpha d\psi dr$. The integration variables range in $r\in (5.5\,\rm kpc, 250\,\rm kpc)$, $\alpha \in (0,\pi)$ and $\psi \in (0, 2\pi)$ with the restriction $\theta \in (0.4^\circ, \theta_c)$, where $\sin\theta = r \sin\phi /(\pmb{r}-\pmb{r}_E)$ (see the lower panel of Fig.~\ref{fig:m31}). To average over the radiation within 100\,kpc around M31, for instance, we have $\theta_c=\rm 100\,kpc/785\,kpc\simeq 7.3^\circ$. The lower limit $0.4^\circ$ for $\theta$ corresponds to a projected size of 5.5\,kpc at a nominal distance of $785\,$kpc, which is the radial extension of M31 viewed by \textit{Fermi}-LAT \citep{Fermi17_M31_excess}. In the analysis of the halo's diffuse gamma-ray emission by \citet{Karwin19}, they remove the emission within $0.4^\circ$ (i.e., the emission of the galaxy itself) so we also exclude it from our calculation as well.

\section{Constraint on the baryon mass in the halo of M31}
\citet{Karwin19} have obtained the average gamma-ray intensity of two annulus regions of $0.4^\circ - 8.5^\circ$ and $8.5^\circ - 14^\circ$ with respect to M31 center, corresponding to, namely, the spherical halo of radius $5.5-120\,$kpc and the far outer halo of radius $120-200\,$kpc respectively. The observed intensities of the two regions above 10 GeV are comparable. However, considering that the CR density and the baryon gas density are lower at larger radius, the expected pionic gamma-ray emissivity in the spherical halo is larger than that in the far outer halo. Therefore, the gamma-ray intensity of the spherical halo is more restrictive and we will perform our analysis based on it. We consider the model-dependent gamma-ray intensity and the model-independent gamma-ray intensity of the spherical halo obtained by \citet{Karwin19} respectively. As mentioned in Section~\ref{sec:intro}, the model-dependent analysis gives a constraining gamma-ray intensity upper limit at 35\,GeV, while the model-independent gamma-ray intensity at the same energy is more conservative. Note that although the projected size of the spherical halo is 120\,kpc, the measured gamma-ray intensity also contains the information of the gas density beyond 120\,kpc because the measured gamma-ray emission results from the superposition of all the emission in the line of sight passing through the entire halo (see Fig.~\ref{fig:m31}). 

\begin{figure}[htbp]
\centering
\includegraphics[width=0.9\columnwidth]{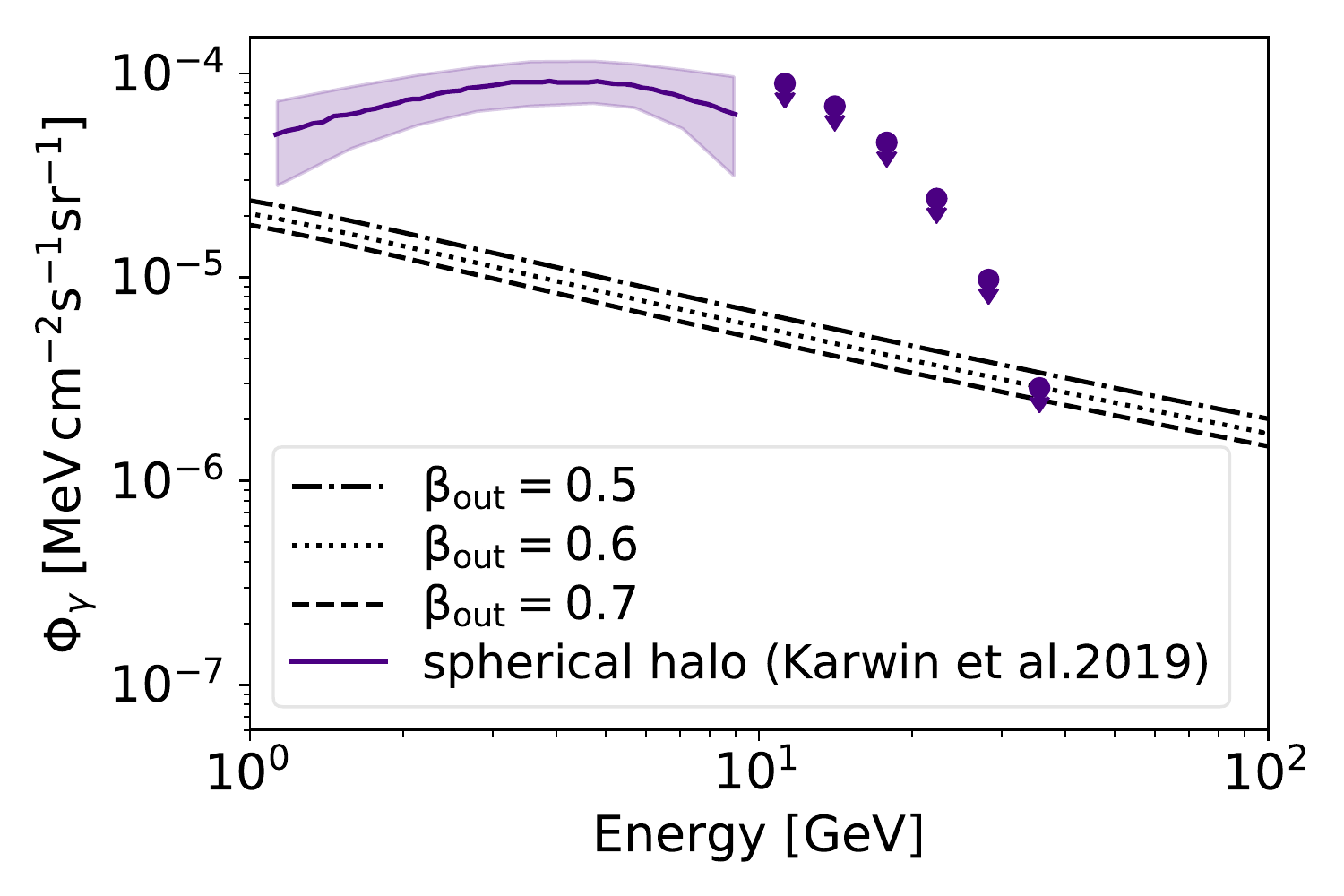}
\includegraphics[width=0.9\columnwidth]{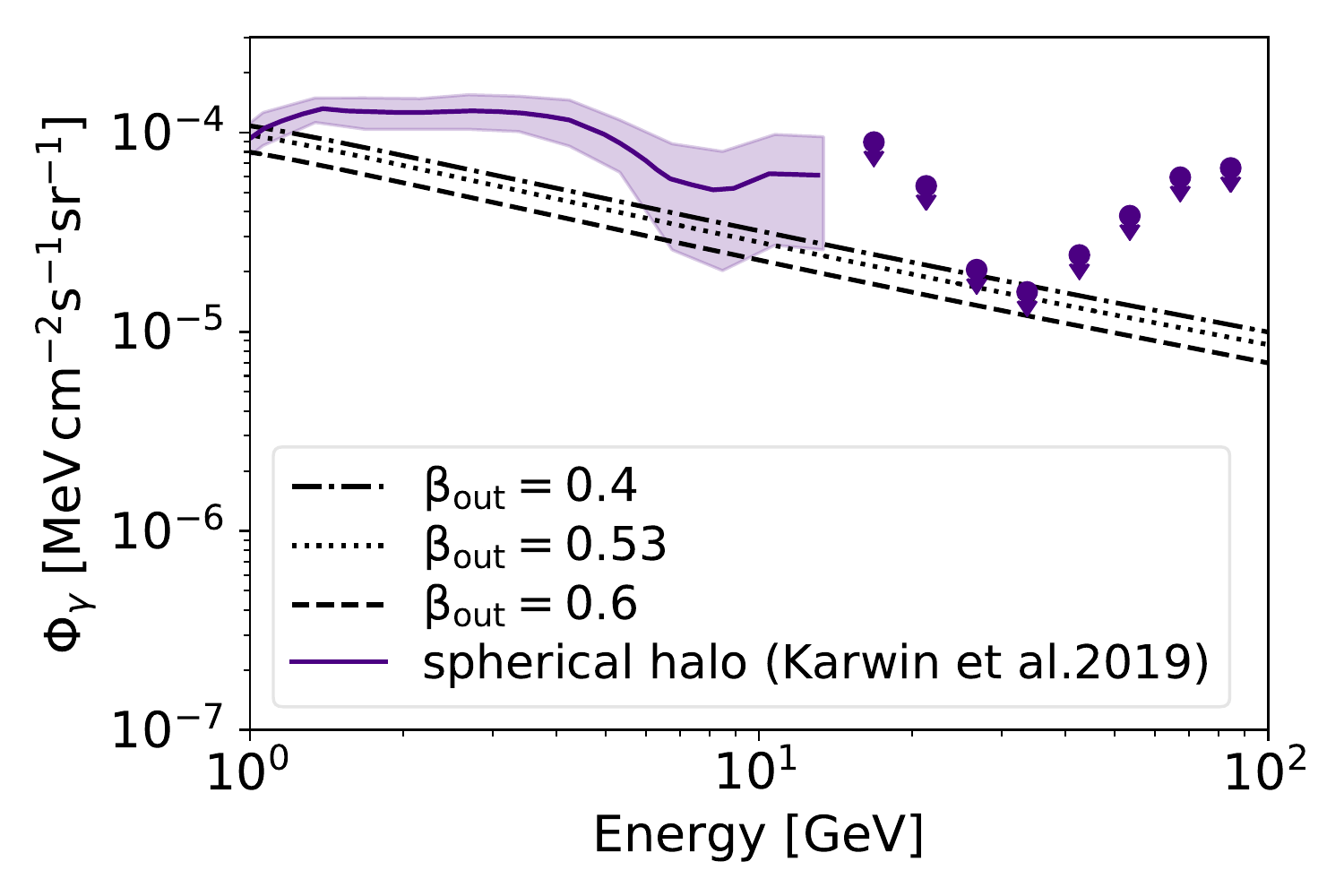}
\caption{Predicted hadronic gamma-ray intensity from the spherical halo with different $\beta_{\rm out}$. The purple band and circles are the observational result given by \citet{Karwin19} for the spherical halo of M31, with the PLEXP model (the top panel) and without model dependence (the bottom panel). We take $s=2.1$, $D_0=6\times10^{30}\,\rm cm^2/s$, $n_0=6.3\times10^{-5}\,\rm cm^{-3}$ and $\beta_{\rm in}=0.6$. }
\label{fig:spec_beta}
\end{figure}

Bearing in mind that the theoretical gamma-ray intensity should not exceed the measured one, we adjust the value of $\beta_{\rm out}$ to search the flattest gas density distribution from 50\,kpc to 250\,kpc allowed by the gamma-ray data. To do this, we scan the entire parameter space of $\beta_{\rm in}$, $\beta_{\rm out}$ and $n_0$ for a given diffusion coefficient. Note that the gamma-ray emissivity is dependent not only on the gas density but also on the CR density. As mentioned earlier, the value of the diffusion coefficient at 350\,GeV (i.e., $D_0$) is crucial to the latter. The standard nonlinear theory for CR diffusion \citep{Yan08} predicts a diffusion coefficient of $10^{29}\rm cm^2s^{-1}$ at several hundreds GeV for the inner halo ($r\sim 10\,$kpc) of our Galaxy, which is comparable to that in the Galactic disk \citep[e.g.][]{Ptuskin06, Strong10, Genolini15, Huang20}. The diffusion coefficient in the outer halo ($r\sim 100\,$kpc) is probably different given different environmental conditions such as the plasma density and the magnetic field strength. From an energetic point of view, the turbulence in the halo might be advected from the galactic disk, or might be driven locally by the differential rotation of the CGM of M31 as observed by XMM-Newton \citep{HK16} and Planck \citep{Tahir19}, or the proper motion of satellite galaxies in the halo of the galaxy with a velocity of a few $\times 100\rm km/s$ \citep{vanderMarel19, Grcevich09,  Watkins10}. A further investigation on the property of the turbulence in M31's halo is beyond the scope of this work, but we note that in some previous literature, the authors generally considered a roughly $10-100$ times larger diffusion coefficient for the halo than that of the Galactic disk \citep{Feldmann13, Kalashev16, Liu19_CGM}. We follow this treatment and consider $D_0$ ranging from $10^{29}\rm cm^2s^{-1}$ up to $10^{31}\rm cm^2s^{-1}$ and caveat that a larger $D_0$ might be also possible if the turbulence in the halo is very weak and it would relax the obtained constraint. In a recent study, the Feedback In Realistic Environments (FIRE) simulation of CR transport in galaxy ISM and CGM shows that for the constant-diffusivity model, the observational constraints of our Galaxy and nearby starburst galaxies could be consistent with a diffusion coefficient $(3-30)\times 10^{29}\rm cm^2s^{-1}$ for 1\,GeV CRs \citep{Chan19,Hopkins21b}. Extrapolating the diffusion coefficient to 350\,GeV with $D(E)\propto E^{1/3}$, one may get a diffusion coefficient $D_0=(2-20)\times 10^{30}\rm cm^2s^{-1}$ at 350\,GeV. On the other hand, \citet{Yan08} considered dominance of the magnetohydrodynamic fast modes in the inner halo of our Galaxy and found an almost energy-independent $D(E)$ in 1\,GeV--1\,TeV, resulting in $D_0=(3-30)\times 10^{29}\rm cm^2s^{-1}$ if we extrapolate $D(E)$ to 350\,GeV accordingly. Both of the extrapolated diffusion coefficient are covered by the chosen range for $D_0$ here.


\begin{deluxetable}{cccccc} 
\tabletypesize{\scriptsize}
\tablewidth{0pt} 
\tablenum{1}
\tablecaption{Constraints on the $\beta_{\rm out}$ for gas distribution beyond 50\,kpc and the total baryon mass enclosed in the halo.\label{tab:beta}}
\tablehead{
\colhead{}&\colhead{$D_0$ ($\rm cm^2/s$)} & \multicolumn{2}{c}{$M_{< 50\,{\rm  kpc}}=5\times 10^9 M_\odot$}& \multicolumn{2}{c}{$M_{< 50\,{\rm  kpc}}=2\times 10^9 M_\odot$} \cr
\cline{3-6}
\colhead{} & \colhead{} & \colhead{$\beta_{\rm out}$} & \colhead{$M_{< 250\,\rm kpc}$ ($M_\odot$)} & \colhead{$\beta_{\rm out}$} & \colhead{$M_{< 250\,\rm kpc}$ ($M_\odot$)}
} 

\startdata 
&$2\times 10^{30}$&$\infty$&$<2\times10^9$&$\infty$&$<2\times10^9$\\
Case&$3\times 10^{30}$&$\infty$&$2.5\times10^9$&2&$2.8\times10^9$\\
1&$6\times 10^{30}$&2&$6.9\times10^{9}$&0.6&$1.4\times10^{10}$\\
&$10^{31}$&0.8&$2.1\times10^{10}$&0.27&$3.9\times10^{10}$\\
\cline{1-6}
&$1\times 10^{30}$&$\infty$&$<2\times10^9$&1.7&$3\times10^9$\\
Case&$2\times 10^{30}$&1.7&$7.8\times10^9$&0.53&$1.7\times10^{10}$\\
2&$3\times 10^{30}$&0.83&$1.95\times10^{10}$&0.27&$3.9\times10^{10}$\\
&$4\times 10^{30}$&0.53&$4.06\times10^{10}$&0.13&$6.3\times10^{10}$\\
\enddata

\tablecomments{The second column is the employed diffusion coefficient for 350\,GeV CR protons. The third and fourth column show the largest values of $\beta$ for the gas density profile outside 50\,kpc and the corresponding baryon mass within 250\,kpc allowed by gamma-ray observation under different diffusion coefficients with the assumption of $M_{< 50\,{\rm  kpc}}=5\times 10^9 M_\odot$. The fifth and sixth column are the same but with the assumption of $M_{< 50\,{\rm  kpc}}=2\times 10^9 M_\odot$.}
\label{tablegas}
\end{deluxetable}

\begin{figure}[htbp]
\centering
\includegraphics[width=0.9\columnwidth]{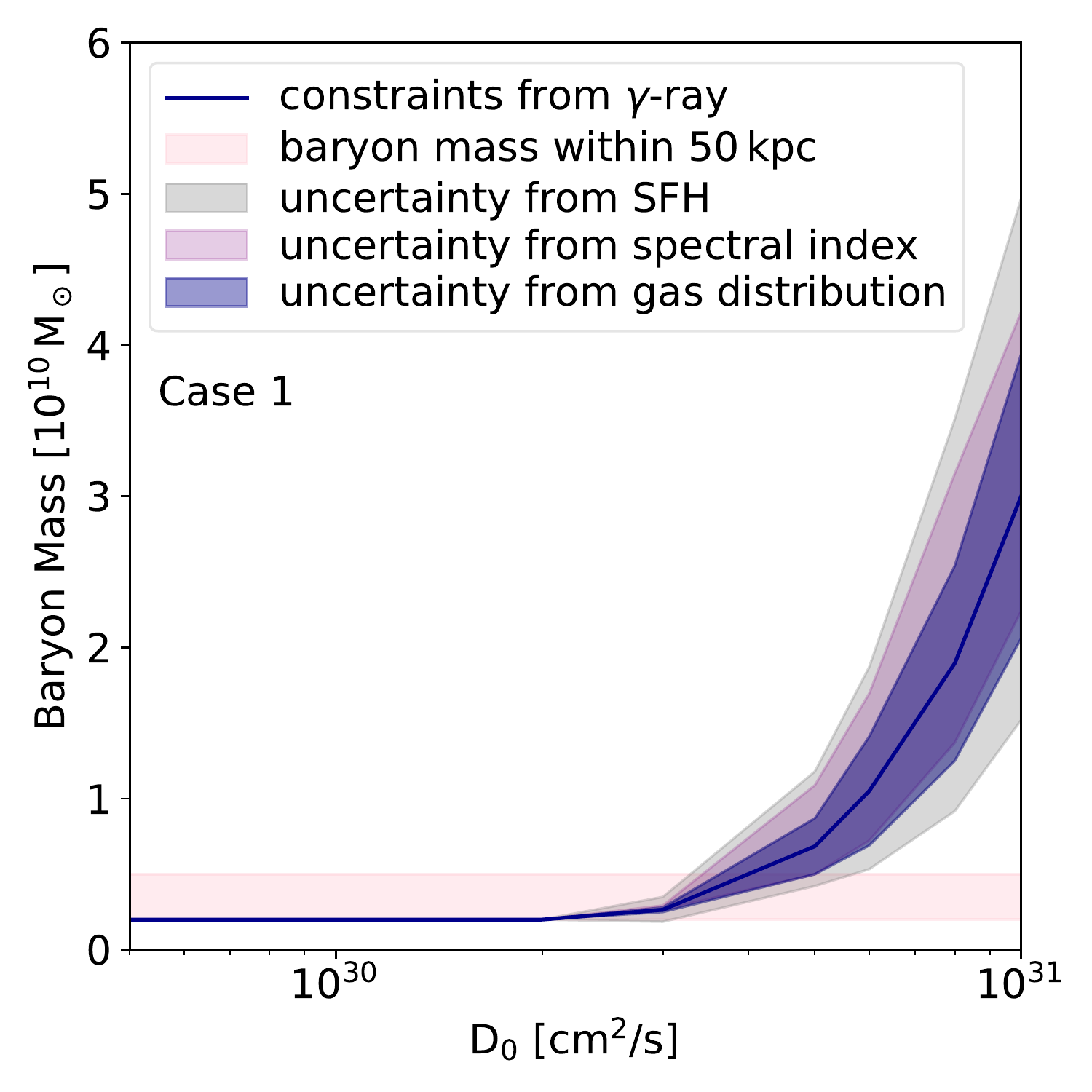}
\includegraphics[width=0.9\columnwidth]{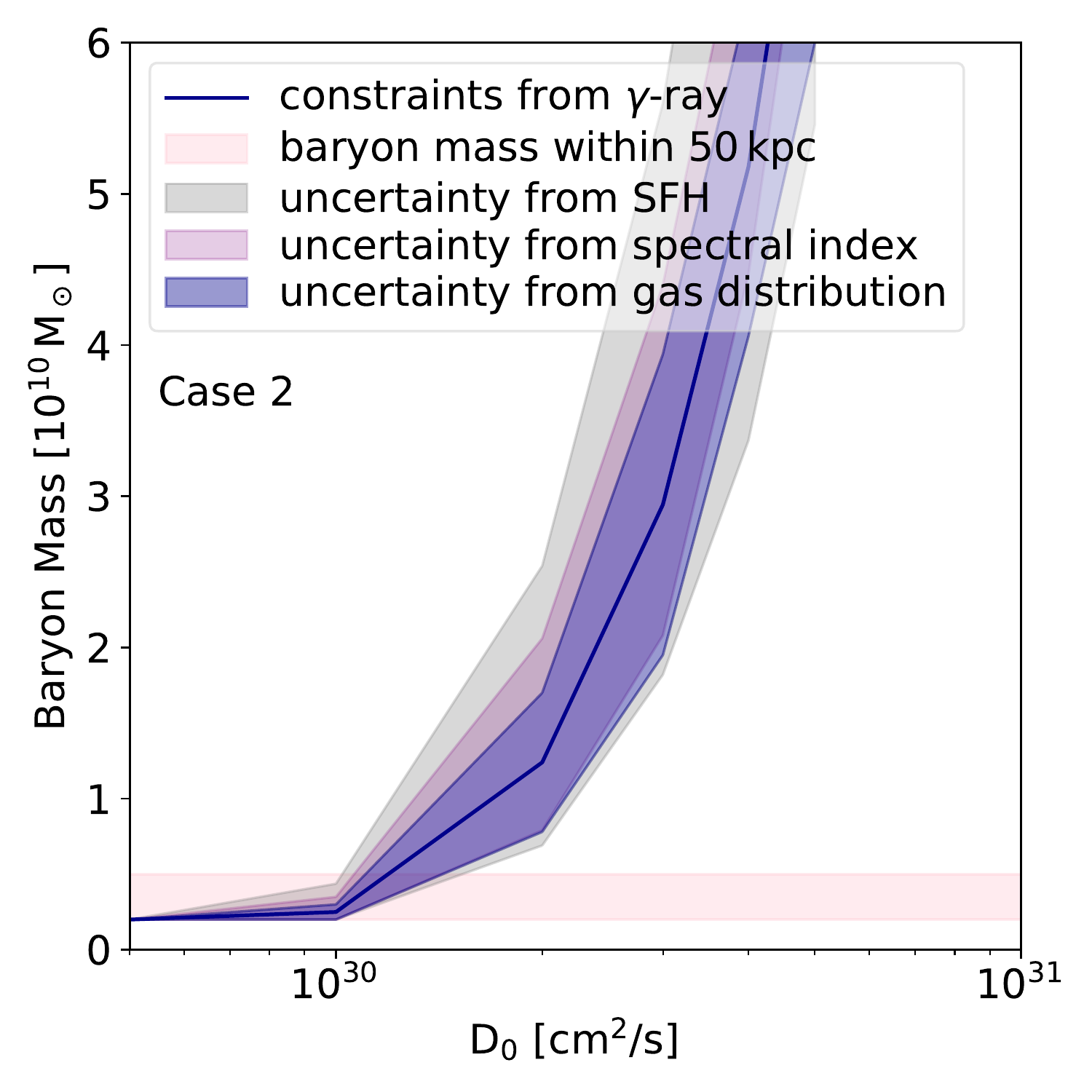}
\caption{Constraint from gamma-ray observation on the mass of the missing baryon within related to M31 that can be explained by CGM within a radius of 250\,kpc. The top (bottom) panel is the result of Case 1 (Case 2). The blue band shows the influence of uncertainties in the model parameters of the gas distribution within 50\,kpc, while the horizontal pink band shows the corresponding baron mass within 50\,kpc. The purple band represents the uncertainties caused by the CR injection spectral index. The gray band shows the uncertainties caused by SFH. The blue solid curve shows the median value of the constraint.}
\label{fig:constraint}
\end{figure}

The critical value of $\beta_{\rm out}$ is obtained when the theoretical gamma-ray intensity touches the lowest data point (i.e., the data point at $35\,$GeV) in the $E_\gamma-\Phi_\gamma$ plot. The obtained $\beta_{\rm out}$ corresponds to the maximum allowable baryon mass in the halo for each $D_0$ considered. In Fig.~\ref{fig:spec_beta}, we show an example that how the predicted gamma-ray intensity changes with the value of $\beta_{\rm out}$. We note that the gamma-ray intensity at $\sim 1$\,GeV may be more constraining especially for a soft injection spectrum. However, the streaming instability might operate for low energy CRs, the density of which is comparatively high \citep{Skilling71, Yan02, Farmer04}. As a result, these low-energy CRs could be scattered by a self-generated waves with reduction of the diffusion coefficient but on the other hand suffering from adiabatic cooling \citep{Pakmor16}. Therefore, the resulting gamma-ray intensity at 1\,GeV may be subject to large uncertainty and hence we do not use it.

We first take the gamma-ray intensity upper limit of the PLEXP model at $\approx 35$\,GeV to constrain the baryon mass (which is referred to as Case 1). As listed in Table~\ref{tab:beta}, for $D_0$ smaller than $2\times 10^{30}\rm cm^2s^{-1}$, the gamma-ray intensity generated within 50\,kpc already exceeds the data point. This result may imply that the CR diffusion in the halo cannot be too slow and in turn put a constraint on the diffusion coefficient (and consequently the properties of the turbulence) in the halo. A flattening of the density profile beyond 50\,kpc would be allowed only if a large diffusion coefficient $\gtrsim 6\times10^{30}\rm cm^2s^{-1}$ for the halo is employed. By translating the value of $\beta_{\rm out}$ to the mass of CGM, we find that the baryon mass of CGM, as presented in Table~\ref{tab:beta} and the top panel of Fig.~\ref{fig:constraint},  cannot exceed $M_{\rm b}\leq (1.4-5)\times 10^{10}M_\odot$ baryons even with a quite large diffusion coefficient $D_0=10^{31}\,\rm cm^2s^{-1}$.

We also repeat the above process with the model-independent upper limit of the gamma-ray intensity at $\approx 35$\,GeV (which is referred to as Case 2). The obtained results are shown in Table~\ref{tab:beta} and the bottom panel of Fig.~\ref{fig:constraint}. Due to the upper limit of gamma-ray intensity used in Case 2 is five times higher than that in Case 1, the constraint on the baryon mass of CGM obtained in Case 2 is more relaxed. The constrained baryon mass would exceed $6\times10^{10}M_\odot$ if $D_0>4\times10^{30}\,\rm cm^2/s$, and then increase very rapidly with increasing $D_0$.
 
The baryon fraction depends on the total mass of the dark matter halo and the total stellar mass. The total mass of the dark matter halo has been estimated by various groups with different methods and the results are subject to variation of a factor of a few. Generally, the total mass of M31's dark matter halo is likely in the range of $M_h=(0.6-2.4)\times 10^{12}\,M_\odot$ \citep[e.g.][]{LiWhite08, Watkins10, Corbelli10, vanderMarel12, Tamm12, Fardal13, Phelps13, Diaz14, Veljanoski14, Sofue15, Carlesi17, Patel17, Kafle18, Fire19, Hestia20, Lemos20}, while the total stellar mass in Andromeda is about $M_*=(1-1.5)\times 10^{11}M_\odot$ \citep{Tamm12, Williams17}. If we take the median value of all the quantities, we will arrive at that the CGM can only explain no more than 30\% of the missing baryon for $D_0\leq 10^{31}\,\rm cm^2s^{-1}$ in Case 1 and no more than $50\%$ for $D_0\leq 4\times10^{30}\rm cm^2s^{-1}$ in Case 2. The result of Case 1 is consistent with the studies on the baryon fraction in Milky Way's halo by various authors\citep{Miller15, Salem15, LiJT17}. If this conclusion can be generalized to other galaxies, it implies that a considerable fraction of the missing baryons probably resides somewhere beyond the virial radius of galaxies, such as in the filaments of the cosmic web \citep{Eckert15,deGraaff19} or in the intracluster medium of galaxy groups \citep{Lim20}. However, due to the uncertainties of the estimation of $M_h$, $M_*$ as well as the gas distribution within 50\,kpc, the maximum percentage of missing baryons that the CGM of M31 can account for, constrained from the gamma-ray observation, ranges in $(5-100)\%$. This prevents us from drawing a robust conclusion. Note that the first two uncertainties are inevitable regardless of the approach that is taken to measure the baryon fraction. Deeper observations of the halo region of M31, particularly the spectroscopic measurement \citep{Yan07}, and better understanding on CR diffusion in CGM are needed to constrain the baryon mass of CGM more accurately.

There may be interesting implication by confronting the baryon mass constrained in this work with the result in other studies. For example, \citet{Lehner20} obtained the mass of cool and warm gas of M31 within 230\,kpc to be $7.2\times10^{9}M_\odot(Z/Z_\odot)^{-1}$. If we compar this result with the upper limit of the total baryon mass, e.g., $5\times10^{10}M_\odot$ in Case 1 for $D_0=10^{31}\rm cm^2s^{-1}$, it would require the metallicity $Z\geq 0.14\,Z_\odot$ in the halo. On the other hand, if the metallicity in the extended halo can be better determined in the future, it would shed light on the total baryon budget in the halo. We note that Fig.~\ref{fig:constraint} could be also regarded as a constraint on the diffusion coefficient or the property of the turbulence in the extended halo. Provided that a better measurement on the baryon budget in M31 with other methods, it then implies that the diffusion coefficient cannot be too small or the turbulence cannot be too strong in the halo. 

\section{Summary}
To summarize, we attempted to utilize gamma-ray observations as a novel and independent probe of missing baryons in this work. Different from some traditional measurements through UV/X-ray observation which are sensitive to gas in certain specific temperature range, the hadronic gamma-ray flux is sensitive to baryonic gases in all phases and does not rely on their chemical composition. We pointed out that the constraint obtained with this method is dependent on the CR diffusion coefficient in the extended halo of the galaxy which is not clearly known at present. Considering a wide range of the diffusion coefficient (up to two orders of magnitude larger than the one of the Galactic plane), we arrived at that the total mass of baryons concealed in the halo of the Andromeda Galaxy could be at most $(1.4-5)\times 10^{10}M_\odot$ for $D_0\leq 10^{31}M_\odot$ (for 350\,GeV cosmic rays)  based on the model-dependent analysis result of the gamma-ray intensity in the extended halo of the galaxy (Case 1), while no more than $6\times 10^{10}M_\odot$ for $D_0\leq 4\times 10^{30}M_\odot$ based on the model-independent analysis result (Case 2). In the latter case, the method becomes unable to constrain the baryon mass with the present gamma-ray data for a larger $D_0$. A flattening of the average CGM density profile beyond 50\,kpc is allowed if a large diffusion coefficient in the halo $D_0>6\times 10^{30}\rm cm^2/s$ presents for Case 1 and $D_0>2\times 10^{30}\rm cm^2/s$ for Case 2. The missing baryon fraction that can be accounted for by the CGM also depends on the total mass of the dark matter halo of Andromeda Galaxy and the total stellar mass in the galaxy, both of which are subject to uncertainties. Employing median values of for these relevant quantities, we would get that the fraction is about 30\% at most for Case 1. But using the more conservative gamma-ray intensity upper limit (Case 2) and uncertainties of other parameters such as $D_0$ lead us to the conclusion that CGM may explain less than $(5-100)\%$ of the missing baryons related to the galaxy. To get a more accurate estimation of the missing baryon fraction in the CGM, we need a better understanding on the properties of the turbulence in the extended galactic halo, a more accurate measurement on the total mass of the dark matter halo and stellar mass of Andromeda Galaxy. In addition, the next generation gamma-ray detectors may perform advanced observation on the diffuse gamma-ray flux or deeper upper limits of the halo of nearby galaxies, making gamma-ray emission a more accurate probe of the missing baryons of galaxies.

\section*{Acknowledgements}
We would like to thank Zhijie Qu for useful comments. This work is supported by NSFC grants 11625312, 11851304 and U2031105, and the National Key R \& D program of China under the grant 2018YFA0404203. HL was supported by NASA through the NASA Hubble Fellowship grant HST-HF2-51438.001-A awarded by the Space Telescope Science Institute, which is operated by the Association of Universities for Research in Astronomy, Incorporated, under NASA contract NAS5-26555. 

\bibliographystyle{apj}
\bibliography{ms210205}



\end{document}